# Generating a hexagonal lattice wave-field with a gradient basis structure

Manish Kumar[*] and Joby Joseph

*Photonics Research Laboratory, Department of Physics, Indian Institute of Technology Delhi, New Delhi, India - 110016*
*Corresponding author: manishk.iitd@gmail.com*

We present a new, single step approach for generating a hexagonal lattice wave-field with a gradient local basis structure. We incorporate this by coherently superposing two (or more) hexagonal lattice wave-fields which differ in their basis structures. The basis of the resultant lattice wave-field is highly dependent on the relative strengths of constituent wave-fields and a desired spatial modulation of basis structure is thus obtained by controlling the spatial modulation of relative strengths of constituent wave-fields. The experimental realization of gradient lattice is achieved by using a phase only spatial light modulator (SLM) in an optical 4f Fourier filter setup where the SLM is displayed with numerically calculated gradient phase mask. The presented method is wavelength independent and is completely scalable making it very promising for micro-fabrication of corresponding structures.

A crystal structure is made of a lattice where each point of lattice is mapped with a particular basis structure [1]. A large area single step fabrication of periodic (or quasi-periodic) crystal structure is easily realized by multiple beam interference method [2-4]. This method leads to a uniform basis throughout the interference volume. However, some recent advances with spatial light modulator assisted phase engineering approach have led to some unconventional 2D lattice wave-fields. For example, experimental realization of periodic lattice with embedded defect sites [5-7] and generation of some dual-lattice wave-fields [8] have been made possible. The main advantage of this SLM assisted approach is that it is a digitally reconfigurable approach and can achieve a single step large area fabrication of corresponding structures in any light sensitive media.

In present work, we extend the idea of coherent superposition of non-diffracting wave-fields [5-7] to propose the synthesis of a gradient optical phase mask which can be utilized for single step large area fabrication of a hexagonal lattice with a gradient basis structure. We do this by coherently superposing two (or three) periodic hexagonal lattice wave-fields with different basis structures. A weighted subtraction (or sum with a phase difference of π) of these wave-fields gives a lattice wave-field where the local basis structure changes in accordance with the local variation of the relative strengths of constituent wave-fields. As an example, we generate a lattice wave-field with a spatial modulation such that it gradually changes from a hexagonal vortex lattice wave-field, to a simple hexagonal lattice wave-field and then to a honeycomb lattice wave-field. Thus, a single step fabrication of such complex gradient structures is now possible which otherwise could have been achieved only by slow serial writing processes of e-beam lithography [9] or direct laser writing [10]. Such structures could be useful for forming graded index photonic elements which find many applications [11, 12].

The periodic lattice wave-fields considered in this paper are formed by interference of multiple plane-waves. The wave-vectors of these constituent plane-waves are spread symmetrically along the surface of a cone with half angle $\theta$ where the axis of the cone is along the vector direction normal to the interference plane. Hence, the parallel component of each of these wave-vectors is equal leading to no modulation along the wave propagation direction i.e. the generated wave-fields are non-diffracting in nature [13]. Since these wave-fields are easily obtained by a sum of multiple plane waves with predesigned initial phase offsets, they may be mathematically represented by Eq. (1) where $n$ is the total number of plane waves interfering together. $\mathbf{E}_j$, $\mathbf{k}_j$ and $\varphi_j$ represent the field amplitude (with polarization), wave-vector and initial phase offset associated with $j^{th}$ plane wave respectively. Moreover, the wave-vector $\mathbf{k}_j$ could be represented in terms of angle $\theta$ by Eq. (2) where $k = 2\pi/\lambda$ ($\lambda$ being the wavelength of laser in air) and coefficient $q_j = 2(j-1)/n$.

$$\mathbf{E}_{n\_\text{beam}}(\mathbf{r}) = \sum_{j=1}^{n} \mathbf{E}_j e^{i(\mathbf{k}_j \cdot \mathbf{r} + \varphi_j)} \quad (1)$$

$$\mathbf{k}_j = k \times [\cos(q_j\pi) \times \sin\theta, \sin(q_j\pi) \times \sin\theta, \cos\theta] \quad (2)$$

Table 1. Synthesis of various periodic lattice wave-fields

| Wave-field | $n$ | $\varphi_j$ |
|---|---|---|
| $\mathbf{E}_{6Ph}$ | 6 | 2π/6, 4π/6, 6π/6, 8π/6, 10π/6, 12π/6 |
| $\mathbf{E}_{3Ph}$ | 3 | 2π/6, 6π/6, 10π/6 |
| $\mathbf{E}_6$ | 6 | 0, 0, 0, 0, 0, 0 |
| $\mathbf{E}_3$ | 3 | 0, 0, 0 |

In this letter, we have considered four lattice wave-fields which are made by superposition of multiple plane waves of same polarization state and making same tilt angle $\theta$. We have taken 0.2° as the value of $\theta$ for all the wave-fields considered in this paper. Only parameters changing for the wave-fields are, number of plane waves i.e. $n$ and initial phase offsets of each of these plane waves i.e. $\varphi_j$. Table 1 summarizes the construction of these wave-fields. We calculate the fields $\mathbf{E}_{6Ph}$ and $\mathbf{E}_{3Ph}$, and show them in Fig. 1. Ph in the symbols $\mathbf{E}_{6Ph}$ and $\mathbf{E}_{3Ph}$ indicates that these wave-field are formed by specially designed wave-vectors with non-zero initial phase offsets as per the values given in Table 1. We have used a selective coloring method to represent the complex fields where color represents the local phase (as mapped by the corresponding colorbar) and the brightness of color is proportional to the local amplitude of the field. The inset images in first row of Fig. 1 represent numerically calculated Fourier transform (FT) of corresponding wave-fields and second row shows corresponding intensity profiles. We also calculate ($E_{6Ph}$-$E_{3Ph}$) and it is clear from Fig. 1(c1) and 1(c2) that this results into a lattice wave-field resembling a simple 3 beam interference field. While comparing Fig. 1(a2) with Fig. 1(c2) we see that the overall lattice remains hexagonal but the basis structure has changed (from a hexagonal vortex lattice to simple hexagonal lattice) owing to the coherent subtraction of $E_{6Ph}$ and $E_{3Ph}$ wave-fields. If this interference pattern were recorded in a photoresist medium then this would lead to a change in the fill-factor of the recorded structure.

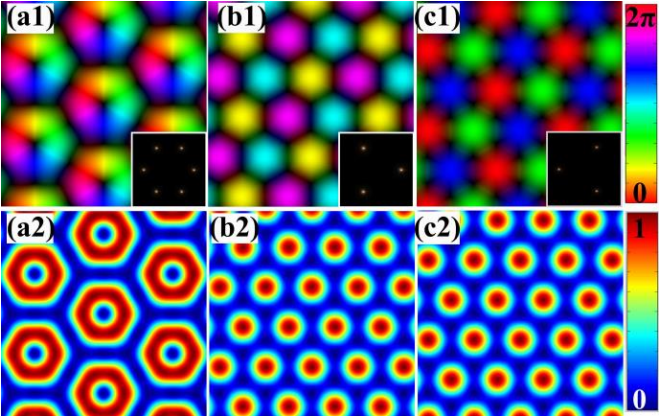

Fig. 1: (Color online) Numerically calculated normalized complex wave-field and corresponding intensity pattern for: $E_{6Ph}$ in (a1) & (a2); for $E_{3Ph}$ in (b1) & (b2); and for ($E_{6Ph}$-$E_{3Ph}$) in (c1) & (c2) respectively. The inset images are numerically calculated FT of corresponding complex wave-fields.

It would be very interesting if this transition between hexagonal vortex lattice wave-field to simple hexagonal lattice field could be controllable. We notice that to obtain the results of Fig. 1(c1) and 1(c2), the two wave-fields were subtracted with full strengths. Next, we see what happens if different strengths of the constituent wave-fields are used. For this we represent the resultant wave-field by a simple expression of Eq. 3 where $a$ controls the relative strength of the constituent wave-fields in this coherent subtraction.

$$\mathbf{E}_{Res} = \mathbf{E}_{6Ph} + a \times \mathbf{E}_{3Ph} \times e^{i\pi} \quad (3)$$

Fig. 2. summarizes the results obtained for values of $a$ kept as 0, 0.5 and 1 respectively. It is conclusive from the results that, as the strength of subtracting wave-field changes from 0 to 1 it leads to gradual change of resultant wave-field from a hexagonal vortex lattice to a simple hexagonal lattice wave-field. This is nicely supported by corresponding simulated FT picture as well. We see how the common wave-vectors of both the constituent wave-fields get subtracted and reduce in their intensity as $a$ varies from 0 to 1.

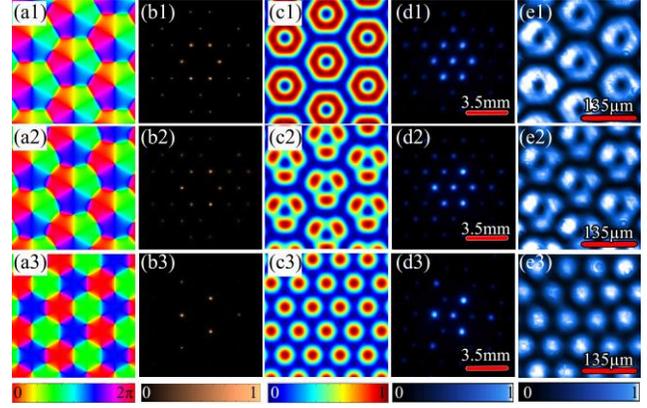

Fig. 2: (Color online) Effect of varying strength of $a$ on resultant wave-field. $a$ = 0, 0.5 and 1 for row number 1, 2 and 3 respectively. Phase profile of $E_{Res}$ in (a1)-(a3); numerically calculated FT of corresponding phase profile of first column in (b1)-(b3); numerically simulated interference pattern in (c1)-(c3); experimentally recorded FT intensity profile in (d1)-(d3); and experimentally obtained interference pattern in (e1)-(e3).

The wave-fields involved here belong to a class of discrete non-diffracting wave-field obtained through superposition of multiple plane-waves. Such non-diffracting wave-fields can be encoded by making use of the phase only filter which is identical with phase only component of the required complex wave-field [6, 7, 14, 15]. This is possible because the error introduced due to phase only approximation for such wave-fields appear in higher order diffraction terms in the Fourier plane which can easily be removed by making use of a Fourier filter mask. Therefore, experimental realization of the required wave-fields was achieved by displaying the extracted phase on a phase only spatial light modulator in a 4f optical Fourier filtering setup and employing appropriate filter at the Fourier plane.

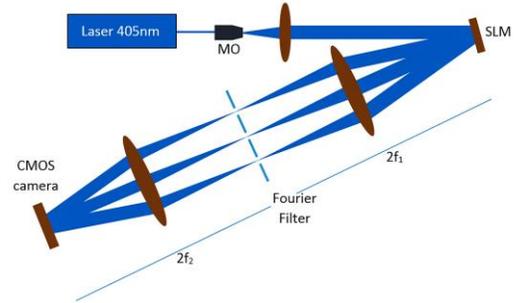

Fig. 3: (Color online) Schematic diagram of the experimental setup. MO: microscope objective

We use a diode laser (Toptica BlueMode, Germany) emitting at 405nm wavelength, a 20x microscope objective and lenses with focal lengths of 135 mm (for collimation), f1 = f2 = 500 mm (for 4f Fourier filter setup), a phase-only SLM (Holoeye-LETO, Germany), and a CMOS camera (DMK-2BUC02, Imaging Source, Germany). Fig. 3 shows the schematic of experimental setup. This setup was used to obtain the results shown in the last two columns of Fig. 2 which have the FT pattern and interference pattern obtained due to use of corresponding phase profile of first column respectively. Their resemblance with corresponding simulation results shown in second and third columns is quite evident. The experimental FT pattern has got a central spot due to the non-zero reflection at SLM-air interface. Since the value of $\theta$ and focal length of FT lens i.e. $f_1$ are known, it is simple to estimate the expected distance between zero and first order diffraction terms in the FT plane as $f_1 \times \sin\theta$ = 500mm × sin(0.2°) ~ 1.745mm. As clear from the scale-bars, this number is totally consistent with experimentally obtained FT pattern of Fig. 2(d1)-(d3).

Having seen the effect of controlled coherent subtraction of two wave-fields on the resultant interference pattern we now take up other two remaining wave-fields from Table-1, namely $\mathbf{E}_6$ and $\mathbf{E}_3$, and do a similar coherent subtraction. Results are summarized in Fig. 4 where the complex wave-fields and the corresponding extracted phase patterns are shown. Comparison of Fig. 1(c1) and Fig. 4(c1) shows a very striking similarity of the corresponding complex wave-fields namely, ($\mathbf{E}_{6Ph}$-$\mathbf{E}_{3Ph}$) and ($\mathbf{E}_6$-$\mathbf{E}_3$). These wave-fields are same except for a transverse shift of ($P_h/2$, $P_h/(2\sqrt{3})$) where $P_h$ is periodicity of the corresponding hexagonal lattice and its value is given by the expression $2\lambda/(3\sin\theta)$. For all the wave-fields in present work, the value of $P_h$ = 2×405nm/(3×sin(0.2°)) ~ 77.35µm. This again is consistent with experimentally obtained results in Fig 2(e1)-(e3) where scale-bars roughly indicate twice the period.

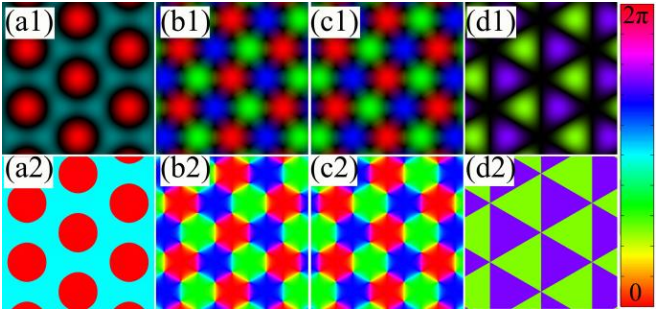

Fig. 4: (Color online) Numerically calculated normalized complex wave-fields and corresponding phase profiles for: $\mathbf{E}_6$ in (a1) & (a2); $\mathbf{E}_3$ in (b1) & (b2); $\mathbf{E}_6$-$\mathbf{E}_3$ in (c1) & (c2); and ($\mathbf{E}_6$-$\mathbf{E}_3$)-$\mathbf{E}_6$/2 in (d1) & (d2) respectively.

We further notice that the intensity profile of the wave-field of Fig. 4(a1) looks like a hexagonal lattice with twice the period of one obtained in Fig. 4(c1), so it could be interesting to do their coherent subtraction. As shown in Fig. 4(d1), a coherent subtraction of wave-fields ($\mathbf{E}_6$-$\mathbf{E}_3$) and $\mathbf{E}_6/2$ leads to a honeycomb lattice wave-field. So, we combine the results from above two sets of coherent subtraction of similar looking wave-fields and use them together to give a very large change in the basis structure of hexagonal lattice wave-field. The resultant lattice wave-field is expressed by the following Eq. 4, where $\mathbf{E}_{6S}$ is just a spatially shifted form of $\mathbf{E}_6$, in order to align it to other constituent wave-fields, and $a$ and $b$ are two parameters which control the relative strengths of the fields to obtain a transition from hexagonal vortex lattice to hexagonal lattice and then to honeycomb lattice wave-field. Table-2 summarizes the value of parameters of Eq. 4 for obtaining these specific wave-fields.

$$\mathbf{E}_{Res} = \mathbf{E}_{6Ph} + a \times \mathbf{E}_{3Ph} \times e^{i\pi} + b \times \mathbf{E}_{6S} \times e^{i\pi} \qquad (4)$$

Table-2. Parameters for obtaining specific lattice wave-fields

| Lattice wave-field | $a$ | $b$ |
|---|---|---|
| Hexagonal vortex | 0 | 0 |
| Hexagonal | 1 | 0 |
| Honeycomb | 1 | 0.5 |

Eq. 4. provides a very powerful expression to control the basis structure of hexagonal lattice wave-field and the phase extracted from resultant field could be used to generate the corresponding interference profile by making use of the experimental setup of Fig. 3. The pattern thus obtained could be recorded into any photo sensitive medium. In all the above treatments, $a$ and $b$ were considered to be spatially constant factors. But there is something very interesting we achieve by making them a function of position i.e. $a$ is replaced by $a(x,y)$ and $b$ is replaced by $b(x,y)$. By doing so we have introduced a spatial modulation in the relative strengths of the constituent wave-fields of Eq. 4 and thus making it possible to introduce a spatial modulation in the basis structure of resultant lattice wave-field as well. In order to confirm this, we create a very simple spatial variation of the lattice structure which covers all three structures of Table 2 as one goes along positive x-direction. Thus we have $a(x,y)$ and $b(x,y)$ reduced to $a(x)$ and $b(x)$ factors respectively which are varied as shown in Fig. 5.

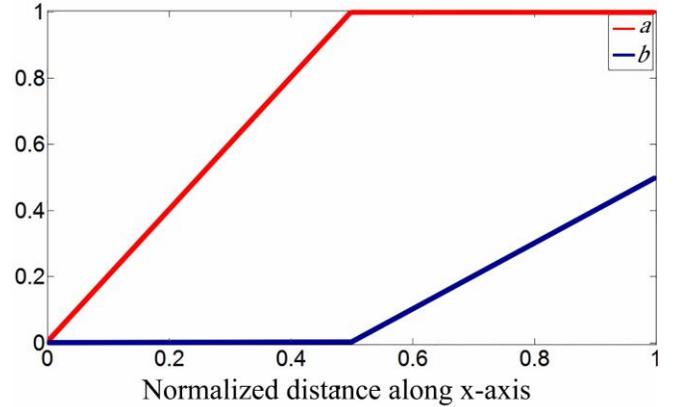

Fig. 5: (Color online) Variation of parameters $a$ and $b$ along x-axis in order to achieve a gradual linear variation of basis structure.

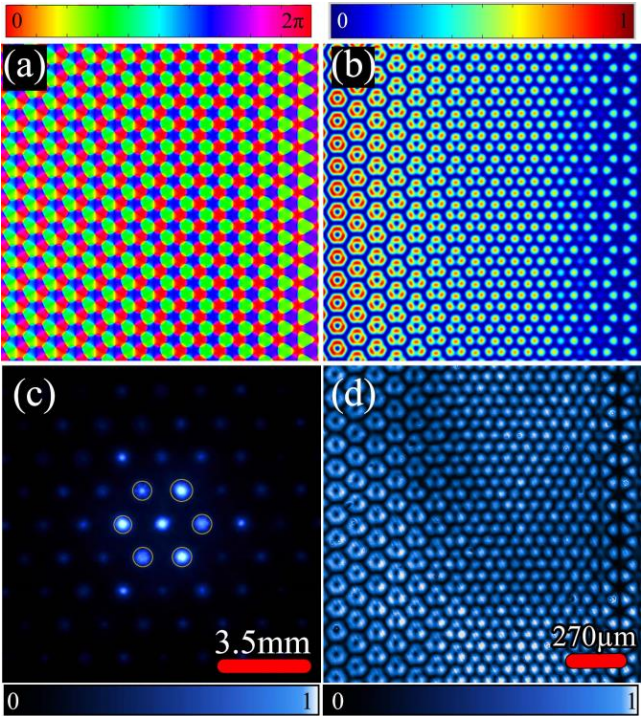

Fig 6: (Color online) Realization of hexagonal lattice wave-field with gradient basis structure. Numerically synthesized gradient phase profile in (a); simulated interference profile in (b); experimentally obtained FT pattern in (c); experimentally obtained interference profile in (d).

Using the spatial variation of $a$ and $b$ as described by the plots of Fig. 5, we obtain $\mathbf{E}_{Res}$. We then extract the phase part of $\mathbf{E}_{Res}$, which is shown in Fig. 6(a). Fig. 6(b) shows the corresponding simulated intensity profile and the gradual variation of local basis structure is very evident from both the profiles. We display the calculated phase profile on the phase only SLM in our experimental setup to obtain the FT profile which is shown in Fig. 6(c). Only the desired plane waves, which are represented by circled spots in the FT plane picture of Fig. 6(c), were allowed to propagate and the rest were blocked using a Fourier filter. The resultant interference pattern as recorded by the CMOS camera is shown in Fig. 6(d) which has a great resemblance with numerically calculated profile of Fig. 6(b). This variation in the basis structure has one to one correspondence with synthesized phase profile/grating. This could be understood in terms of a simple physical argument as follows. Since the diffracted beams are generated according to phase contrast of the grating, a variation in the phase profile leads to a local change in the relative phase offset (and/or diffraction efficiency) of diffracted beams. Thus each of the diffracted beams gets spatially modulated on account of this local change in the phase offset (and/or diffraction efficiency). A recombination of the first order diffracted beams, which are selected through the Fourier filtering process of the experimental setup, then give rise to a change in the local basis structure.

In conclusion, we have proposed a method for calculation of gradient phase mask for generating a lattice wave-field with gradient basis structure. As an example we have worked with hexagonal lattice wave-field and have shown a linear gradient basis structure. Such profiles could easily be recorded into any photosensitive medium. As the method is completely scalable in terms of wavelength of light and also by employing the demagnification factor in the 4-f Fourier filtering setup, it could easily be used for micro-fabrication of corresponding gradient structure to form graded index photonic elements [11, 12]. The generated wave-fields, being a class of vortex array, could be used in particle trapping and manipulations as well [16]. The method could easily be extended to other set of non-diffracting wave-fields to give new gradient basis structure designs. With the addition of one on-axis propagating wave it could be possible to realize a complex 3D gradient structure as well [17]. Next, it is interesting to note that when our method of coherent subtraction with a spatial modulation is applied to a wave-field with its own copy, it leads to an interference pattern with variation in local fill-factor and the corresponding synthesized phase filter in this case is the same as one synthesized by an approach proposed by Davis et al. [18]. Thus our proposed method, of spatially modulated coherent subtraction of non-diffracting wave-fields, gives rise to a generalized phase filter which not only is able to directly control the overall modulation of the interference pattern but also controls relative strengths and relative phase offsets of various constituent wave-vectors leading to more exciting results.